\documentclass[10pt,twocolumn]{article}

\usepackage{amsmath,amssymb}
\usepackage{graphicx}

\newcommand{\be}{\begin{equation}}
\newcommand{\ee}{\end{equation}}
\newcommand{\eq}[1]{(\ref{#1})}

\def\beq{\begin{eqnarray}}
\def\eeq{\end{eqnarray}}\def\beqa{\begin{eqnarray}}
\def\eeqa{\end{eqnarray}}

\def\vq{{\bf q}}

\def\vk{{\bf k}}

\begin{document}
\title{$d$-wave bond-order charge excitations in electron-doped cuprates}

%

\author{Hiroyuki Yamase\\
{\small National Institute for Materials Science, Tsukuba 305-0047, Japan}\\
 Mat\'{\i}as Bejas, Andr\'es Greco\\
{\small Facultad de Ciencias Exactas, Ingenier\'{\i}a y Agrimensura }\\
{\small and Instituto de F\'{\i}sica Rosario (UNR-CONICET),} \\
{\small Avenida Pellegrini 250, 2000 Rosario, Argentina}
}


\date{June 19, 2015}

%
%

\twocolumn[
  \begin{@twocolumnfalse}
    \maketitle
\begin{abstract}
We study charge excitation spectra in the two-dimensional $t$-$J$ model on 
a square lattice 
to explore a charge-order tendency recently found in electron-doped cuprates 
around the carrier density 0.15. The static susceptibility of $d$-wave  charge density, 
which corresponds to the nematic susceptibility at the momentum transfer $\vq=(0,0)$, 
shows two characteristic peaks at momenta of the form 
$\vq_{1}=(q',q')$ and $\vq_{2}=(q,0)$. These two peaks 
originate from the so-called $2k_{F}$ scattering processes enhanced by 
the $d$-wave character of the bond-charge density. The peak at $\vq_{1}$ is much broader, 
but develops to be very sharp 
in the vicinity of its instability, whereas the peak at $\vq_{2}$ 
becomes sharper with decreasing temperature, but does not diverge. 
The equal-time correlation function, which is measured by resonant x-ray scattering, 
exhibits a momentum dependence similar to the static susceptibility. 
We also present energy-resolved charge excitation spectra. The spectra show 
a V-shaped structure around $\vq=(0,0)$ 
and bend back toward close to zero energy due to the charge-order tendency 
at $\vq_{1}$ and $\vq_{2}$. The resulting spectra form gap-like features  
with a maximal gap at $\vq \approx \vq_{1}/2$ and $\vq_{2}/2$. 
We discuss implications for the recent experiments in electron-doped cuprates.  
\end{abstract}

\vspace*{1cm}
{\small
\begin{tabular}{ll}
 PACS: & 74.72.Ek {\it Electron-doped cuprates}; \\
       & 75.25.Dk {\it Orbital, charge, and other orders, including coupling of these orders}; \\
       & 78.70.Ck {\it X-ray scattering}
\end{tabular}
}
\vspace*{1cm}

  \end{@twocolumnfalse}

  ]

\section{Introduction} 
Charge order (CO) in high-temperature cuprate superconductors attracts renewed interest. 
CO is known in La-based materials as a spin-charge stripe order \cite{tranquada95}, 
in which CO is accompanied by a spin order. 
However, a different type of CO has been observed recently in various hole-doped cuprates such as 
Y- \cite{wu11,ghiringhelli12, chang12, achkar12,leboeuf13,blackburn13,blanco-canosa14}, 
Bi- \cite{comin14, da-silva-neto14, hashimoto14}, and Hg-based \cite{tabis14} materials. 
In these materials, the CO is not accompanied by a spin order. 
Furthermore a modulation vector of the CO 
decreases with doping, the opposite tendency observed in the La-based materials. 
The origin of the newly found CO as well as its relation to superconductivity 
is under active debate \cite{meier14,sachdev13,wang14,atkinson15,bejas12}. 

Research interest also goes to electron-doped cuprates. 
Resonant inelastic x-ray scattering (RIXS) reveals that charge excitation spectra 
develop to form a V-shaped dispersion \cite{wslee14} around the momentum 
$\vq=(0,0)$  and extends 
up to around 1.5 eV at $\vq=(0.6 \pi,0)$ and $(0.6 \pi, 0.6\pi)$ 
\cite{ishii14} in Nd$_{2-x}$Ce$_{x}$CuO$_{4}$ with $x=0.15$.  
Quite recently resonant x-ray scattering (RXS), which integrates a RIXS spectrum 
up to infinity with respect to energy, 
has revealed a charge excitation peak at $\vq\approx(0.48\pi,0)$ near $x=0.15$ \cite{da-dilva-neto15}. 
The observed wavevector is rather close to that found in hole-doped cuprates 
\cite{wu11,ghiringhelli12, chang12, achkar12,leboeuf13,blackburn13,blanco-canosa14,comin14, 
da-silva-neto14, hashimoto14, tabis14}, 
implying a possible universal phenomenon for the  CO in cuprate superconductors. 
The correlation length of the CO is, however,  estimated to be 
4-7 lattice constants, i.e., it is not a long-range order\cite{da-dilva-neto15}. 

The wavevector of $\vq\approx (0.48 \pi, 0)$ obtained by RXS \cite{da-dilva-neto15} 
is covered by the RIXS by Ishii {\it et al.} \cite{ishii14}, but the observed RIXS spectra 
do not seem to suggest clearly some characteristic feature associated with a CO, 
such as softening of the spectrum toward a long-range order 
at the corresponding wavevector. This peculiar situation motivates us to study 
charge excitations in electron-doped cuprates more closely from a theoretical point of view.

Charge excitations in electron-doped cuprates are not much known theoretically. 
Ishii {\it et al.} studied the usual density-density  
correlation functions and discussed the RIXS spectra  \cite{ishii14}. 
Bejas {\it et al.} \cite{bejas14}, on the other hand, studied all possible COs 
in the $t$-$J$ model with parameters appropriate for electron-doped cuprates. 
They found that instead of a usual charge-order instability, various types of 
bond order tend to occur much more strongly. 
In particular, a $d$-wave bond-order tendency is dominant in a moderate doping region. 
While its instability is expected at a wavevector close to $(\pi,\pi)$, 
they found a meta-stable solution of the $d$-wave bond order at $\vq \approx (0.49 \pi,0)$. 
This wavevector is very close to the experimental observation by RXS \cite{da-dilva-neto15}. 

Encouraged by this agreement with the experiment, we study charge excitations 
associated with a $d$-wave bond order in the two-dimensional $t$-$J$ model 
on a square lattice
by taking parameters appropriate for electron-doped cuprates. 
We compute three quantities: static $d$-wave bond-order susceptibility $\chi_{d}(\vq,0)$, 
its spectral weight ${\rm Im}\chi_{d}(\vq,\omega)$, and the corresponding 
equal-time correlation function $S(\vq)$. The second and third quantities 
can be measured directly by RIXS and RXS, respectively. 
Our obtained results capture essential features observed in experiments 
such as a V-shape dispersion of ${\rm Im}\chi_{d}(\vq,\omega)$ near $\vq=(0,0)$ \cite{wslee14, ishii14} 
and a short-range CO with $\vq\approx (0.48 \pi, 0)$ \cite{da-dilva-neto15}. 
In addition, we obtain several new insights: 
First, a CO is expected also at $\vq_{1}=(q',q')$ with $q' \approx 0.84\pi$. 
In fact this CO 
has a stronger intensity than the CO at $\vq_{2}=(q,0)$ with $q \approx 0.49\pi$. However, 
the peak at $\vq_1$ is much broader in momentum space than that at 
$\vq_2$ and becomes sharp only in the vicinity of its instability. 
Second, the dispersive peak of ${\rm Im}\chi_{d}(\vq,\omega)$ bends back toward 
close to zero energy at $\vq_{1}$ and $\vq_{2}$ 
where $\chi_{d}(\vq,0)$ and $S(\vq)$ exhibit a peak. 
The resulting charge excitation spectra 
show gap-like features between $\vq_{1}$ and $\vq=(0,0)$, and between  $\vq=(0,0)$ and $\vq_{2}$, with 
a maximal gap at $\vq \approx \frac{1}{2} \vq_{1}$ and  $\frac{1}{2} \vq_{2}$.

\section{Model and formalism}
Various approximations to the $t$-$J$ \cite{gooding94,martins01,bejas14} 
and the strong coupling Hubbard \cite{macridin06} model 
show that the models have a strong tendency toward phase separation, especially for parameters 
appropriate for electron-doped cuprates. 
The phase separation, however, can be an artifact caused by neglecting the 
long-range Coulomb interaction. In fact, the Coulomb interaction term appears naturally 
when the $t$-$J$ model is derived from the three-band Hubbard model \cite{feiner96}.
Hence we include the nearest-neighbor Coulomb interaction in the $t$-$J$ model 
as a minimal model to study electron-doped cuprates. Our model then reads 
\begin{equation}
H = \hspace{-0.1cm }-\hspace{-0.1cm }\sum_{i, j,\sigma} t_{i j}\tilde{c}^\dag_{i\sigma}
\tilde{c}_{j\sigma} + J \sum_{\langle i,j \rangle} \left[ \vec{S}_i \cdot \vec{S}_j-\frac{1}{4} n_i n_j \right]
+V\sum_{\langle i,j \rangle} n_i n_j
\label{H}
\end{equation}
where $t_{i j} = t$ $(t')$ is the hopping between the first (second) nearest-neighbor 
sites on a square lattice, $J$ and $V$ are the exchange and Coulomb interactions between 
the nearest-neighbor sites, respectively.
$\langle i,j \rangle$ indicates a nearest-neighbor pair of sites. 
$\tilde{c}^\dag_{i\sigma}$ ($\tilde{c}_{i\sigma}$) is 
the creation (annihilation) operator of electrons 
with spin $\sigma$ ($\sigma = \downarrow$,$\uparrow$) 
in the Fock space without any double occupancy. 
$n_i=\sum_{\sigma} \tilde{c}^\dag_{i\sigma}\tilde{c}_{i\sigma}$ 
is the electron density operator and $\vec{S}_i$ is the spin operator. 
The role of the $V$-term turns out to merely suppress phase separation in a 
doping and temperature ($T$) range relevant to cuprates. 
In fact, our obtained results are not affected essentially by the $V$-term.

In leading order of a $1/N$-expansion\cite{foussats04}, 
the kinetic term of the electrons is characterized by 
an effective electronic dispersion
\begin{equation} \label{eq:ek}
\varepsilon_{\bf k}= -2 \left[ t \frac{\delta}{2} + J \Delta \right] (\cos k_x+\cos k_y)-4 t'\frac{\delta}{2}\cos k_x \cos k_y - \mu,
\end{equation}
\noindent where $\delta$ is the doping rate away from half-filling and 
$\mu$ is the chemical potential. The bare hopping integrals $t$ and $t'$ are renormalized by a factor of $\delta$.  
The term $J \Delta$ in eq. (\ref{eq:ek}), 
which is not present at bare level, comes from the exchange term (the second term in eq. \eq{H}). 
The magnitude of $\Delta$ describes a bond amplitude between the nearest neighbor sites. 
Values of $\Delta$ and $\mu$ are determined self-consistently at a given $\delta$ 
by solving the following equations: 
\begin{eqnarray}
\Delta=\frac{1}{4 N_s} \sum_{\bf k} (\cos k_x + \cos k_y) n_F(\varepsilon_{\bf k})
\label{Delta}
\end{eqnarray}
and
\begin{eqnarray}
1-\delta=\frac{2}{N_s} \sum_{\bf k} n_F(\varepsilon_{{\bf k}})\,. 
\end{eqnarray}
Here $n_F$ is the Fermi function and $N_s$ the total number of lattice sites.

\begin{figure}
\centering
\includegraphics[width=8cm]{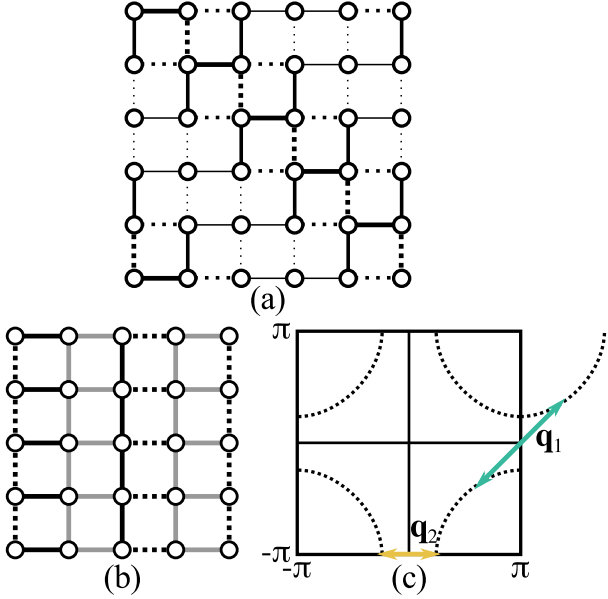}
\caption{(Color online) $d$-wave bond order for $\vq_1 \approx \vq=(0.8\pi,0.8\pi)$ (a) and $\vq_{2}\approx \vq=(0.5\pi,0)$ (b). 
The black lines denote a stronger (solid line) and weaker (dotted line) 
bond relative to the average bond amplitude (gray line), which corresponds to 
$\Delta$ in eq. \eq{Delta}. The width of the lines indicates the modulation amplitude.
(c) $2k_F$ scattering processes, which determine the wavevectors $\vq_{1}$ and $\vq_{2}$. 
}
\label{fig:pattern}
\end{figure}

In the above scheme Bejas {\it et al.} studied all possible charge instabilities for both 
hole-doped\cite{bejas12} and electron-doped\cite{bejas14} cuprates. 
They found that a relevant instability to discuss the electron-doped cuprates around $\delta=0.15$ 
is a $d$-wave bond order where bond amplitude is modulated along 
both $x$ and $y$ direction, and its relative phase is in antiphase \footnote{This order is refereed to 
as BOP$_{x\bar{y}}$ in \cite{bejas14}}. 
This ordering pattern is shown in fig. \ref{fig:pattern} by choosing wavevectors close to those relevant in the present study. 

In the present study, we explore closely a charge-order tendency associated with 
the $d$-wave bond order. In particular, we aim to give some insight into 
the charge excitations recently observed by RIXS and RXS in electron-doped cuprates from a theoretical point of view. 
Following \cite{bejas12} and \cite{bejas14} we focus on the effective dynamical $d$-wave charge susceptibility, 
\begin{eqnarray}
\chi_{d}(\vq, \omega)= \frac{(8J\Delta^2)^{-1}} {1-2J\Pi(\vq, \omega)} 
\label{chi-dPI}
\end{eqnarray}
which becomes exact in leading order of $1/N$. The bare polarizability $\Pi(\vq, \omega)$ reads 
\begin{eqnarray}
\label{Pi-dPI}
\Pi(\vq, \omega) = - \frac{1}{N_{s}}\;
\sum_{\vk}\; \gamma^2(\vk) \frac{n_{F}(\epsilon_{\vk + \vq/2}) 
- n_{F}(\epsilon_{\vk - \vq/2})} 
{\epsilon_{\vk + \vq/2} - \epsilon_{\vk- \vq/2}-\omega - {\rm i}\eta}\,,
\end{eqnarray}
where $\eta(>0)$ is an infinitesimally small value and we take $\eta=10^{-3}$ 
because of a practical reason of numerical computations.  
The form factor 
$\gamma(\vk)=(\cos k_x - \cos k_y)/2$ 
comes from the intra-unit-cell symmetry.  
This form factor has $d$-wave symmetry,  and 
$\chi_{d}(\vq, \omega)$ corresponds to the well known nematic susceptibility 
for $\vq=0$ \cite{yamase00a,yamase00b,metzner00}.  
The property of $\chi_{d}(\vq,\omega)$ near $\vq=0$ was already studied in \cite{yamase04b} 
by focusing on a collective mode of the $d$-wave bond order 
in both paramagnetic and superconducting states. 
Here we study eq. \eq{chi-dPI} in a different situation in which COs tend to occur 
at $\vq=\vq_{1}$ and $\vq_{2}$.

In what follows we present results for $J/t=0.3$ and $t'/t=0.30$, which are 
suitable for electron-doped cuprates. 
We fix the carrier density $\delta=0.15$ so that our results will be compared  
directly with recent experiments \cite{wslee14,ishii14,da-dilva-neto15}.  
Our conclusions do not depend sensitively on a precise choice of parameters and 
we choose $V/t=1$. Below we present all quantity of the dimension of energy in units of $t$. 

\section{Results}
We present the static susceptibility $\chi_{d} (\vq, 0)$, 
the equal-time correlations function $S(\vq)$, and 
the spectral weight ${\rm Im}\chi_{d}(\vq, \omega)$ in $(\vq, \omega)$ space. 
RXS and RIXS measure $S(\vq)$ and ${\rm Im}\chi_{d}(\vq, \omega)$, respectively. 

\begin{figure}
\centering
\includegraphics[width=8.5cm]{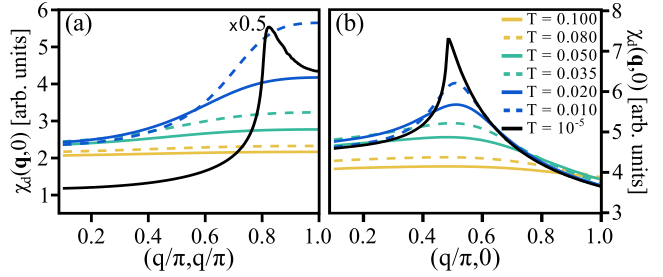}
\caption{(Color online)  The static susceptibility $\chi_d(\vq,0)$ along the directions of $(0,0)$-$(\pi,\pi)$ 
(a) and $(0,0)$-$(\pi,0)$ (b) for various temperatures. 
In (a), the curve at $T=10^{-5}$ is scaled by a factor of 0.5.}
\label{xq}
\end{figure} 

We first computed $\chi_{d} (\vq,0)$ in the entire Brillouin zone and found 
two well-defined peaks 
at $\vq_{1}=(0.84\pi, 0.84\pi)$ and $\vq_{2}=(0.49\pi, 0)$ near zero temperature. 
To clarify their temperature dependence, we plot 
$\chi_{d}(\vq, 0)$ along the $(0,0)$-$(\pi,\pi)$ direction in fig. \ref{xq}~(a). 
At $T=10^{-5}$ a very sharp peak forms at $\vq=\vq_{1}$. 
This peak is due to the proximity to a quantum critical point 
of the $d$-wave bond-order instability, which is present at $\delta_c \approx 0.13$\cite{bejas14}. 
However, once the temperature is increased, the peak is immediately broadened and 
becomes less clear already at $T=0.01$. 
In fig. \ref{xq}~(b) we plot $\chi_{d}(\vq,0)$ along the $(0,0)$-$(\pi,0)$ direction. 
In contrast to fig. \ref{xq}~(a), the temperature dependence of the peak features 
more usual behavior in the sense that the peak is broad at high $T$ and smoothly grows to be 
a pronounced peak at low temperature. In spite of this clear peak structure, 
$\chi_{d}(\vq,0)$ does not diverges at $\vq_{2}$. 
That is, there is no indication that the $d$-wave bond order becomes 
long range along the $(0,0)$-$(\pi,0)$ direction.

\begin{figure}[t]
\centering
\includegraphics[width=8cm]{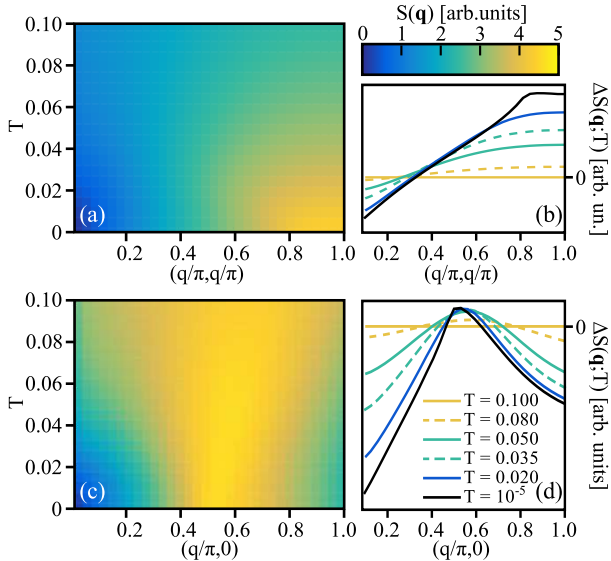}
\caption{(Color online)  (a) Intensity map of $S(\vq)$ along the $(0,0)$-$(\pi,\pi)$ direction for $0<T \leq 0.1$. 
(b) Evolution of the spectral weight $\Delta S(\vq; T)=S(\vq; T) - S(\vq; T=0.1)$  
for various temperatures. Consequently $\Delta S(\vq; T)=0$ at $T=0.1$. 
(c) and (d): corresponding results to (a) and (b), respectively,  
along the $(0,0)$-$(\pi,0)$ direction.}
\label{Sq}
\end{figure} 

The static susceptibility $\chi_{d}(\vq,0)$ is a useful quantity to study 
the stability of a system, i.e., an ordering phenomenon. 
This quantity is, however, not measured directly by RXS. 
Rather, RXS measures the equal-time correlation function $S(\vq)$, which is 
defined by 
\be
S(\vq)=\frac{1}{\pi} \int_{-\infty}^\infty  {\rm d}\omega \, {\rm Im} \chi_{d}(\vq,\omega) 
\left[ n_B(\omega)+1 \right], 
\ee
\noindent where $n_B$ is the Bose factor.
In fig. \ref{Sq}~(a) we show an intensity map of $S(\vq)$ along the $(0,0)$-$(\pi,\pi)$ direction 
in a temperature range $0 < T \leq 0.1$.  
Although the spectral weight tends to accumulate around 
$\vq=\vq_{1}$ with decreasing $T$, the temperature dependence is weak and the 
spectral weight still spreads down to zero temperature in spite of the proximity of the 
corresponding charge instability. 
To show the temperature dependence of $S(\vq)$ more clearly, we plot a spectrum 
$\Delta S(\vq; T)=S(\vq; T) - S(\vq; T=0.1)$ in fig. \ref{Sq}~(b). 
Its temperature  dependence is very similar to that of the static susceptibility shown in 
fig. \ref{xq}~(a) except at $T=10^{-5}$.  
In fig. \ref{Sq}~(c) and (d), we present the corresponding results along the $(0,0)$-$(\pi,0)$ direction. 
Although  the CO tendency is stronger at $\vq=\vq_{1}$ than at $\vq=\vq_{2}$ (see fig. \ref{xq}), 
the peak structure at $\vq=\vq_{2}$ is much clearer, being sharp and 
pronounced with decreasing $T$, as demonstrated in fig. \ref{Sq}~(d).

\begin{figure}[t]
\centering
\includegraphics[width=8cm]{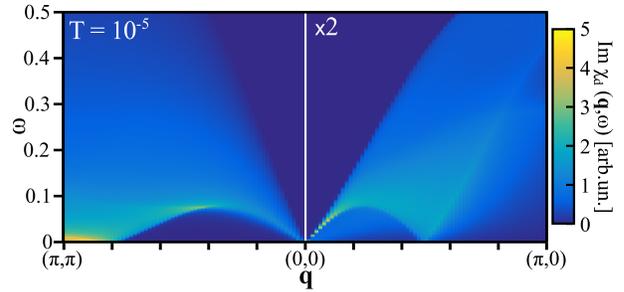}
\caption{(Color online) 
Energy-resolved spectral weight ${\rm Im}\chi_{d}(\vq,\omega)$ in the plane 
of $\vq$ and $\omega$ along the symmetry axes at low temperature.  
The spectral weight is scaled by a factor of 2 along the $(0,0)$-$(\pi,0)$ direction to get 
a better contrast to that along the $(\pi,\pi)$-$(0,0)$ direction.  
}
\label{xqw}
\end{figure} 

The energy-resolved spectral weight ${\rm Im}\chi_{d}(\vq,\omega)$ is shown in  
fig. \ref{xqw} at low temperature in the plane of $\vq$ and $\omega$. 
A V-shape dispersion develops from $\vq=(0,0)$. 
This dispersion originates from individual particle-hole excitations and 
extends up to high energy. 
However, the spectrum bends back 
and softens toward close to zero energy at $\vq=\vq_{1}$ and $\vq_{2}$, where both 
static susceptibility (fig. \ref{xq}) and equal-time correlation function (fig. \ref{Sq}) exhibit a peak. 
These dispersion near $\vq_{1}$ and $\vq_{2}$ is interpreted as coming from 
collective charge excitations. This collective feature is particularly clear near $\vq=\vq_{1}$ 
due to the proximity to the corresponding charge instability. 
A gap-like feature of charge excitations is visible between $\vq_{1}$ and $\vq=(0,0)$, 
and also between $\vq=(0,0)$ and $\vq_{2}$, and forms a maximal gap of about $0.1$
at $\vq \approx \frac{1}{2}\vq_{1}$ and $\frac{1}{2}\vq_{2}$. 
This gap-like feature is more pronounced along the $(0,0)$-$(\pi,\pi)$ direction 
because the $d$-wave character of the bond-charge density 
suppresses its low-energy scattering processes substantially.

\section{Discussions} 
Now we discuss implications for the experiments by RXS\cite{da-dilva-neto15} 
and RIXS \cite{wslee14,ishii14}. 

In fig. \ref{Sq}~(c) and (d), the charge peak at $\vq=\vq_{2}$ becomes sharper with decreasing temperature 
in the equal-time correlation function, but the real part of the susceptibility remains finite 
at the corresponding wavevector (fig. \ref{xq}~(b)). We therefore conclude that the CO 
at $\vq=\vq_{2}$ remains a short range, which is consistent with 
the RXS measurements \cite{da-dilva-neto15}.  
In particular, a short-range feature of the 
observed CO can be interpreted as an intrinsic property, i.e., it does not come from some 
disorders frequently present in actual materials.

On the other hand, the present theory predicts that the CO tendency 
at $\vq=\vq_{1}$ is much stronger than at $\vq=\vq_{2}$. This is because a long-range order with 
modulation vector $\vq \approx  \vq_{1}$ occurs at $T=0$ at the critical 
doping $\delta_c \approx 0.13$ \cite{bejas14}. 
This peak at $\vq_{1}$ is, however, peculiar in the sense that it is much broader than at 
$\vq=\vq_{2}$ and becomes 
sharper only in the vicinity of the charge instability. 
Since $\vq_{1}$ is rather close to $(\pi,\pi)$, it is not straightforward to test it in experiment. 
In fact, it is difficult to perform RXS and RIXS up to near $(\pi,\pi)$.  
Hence a usual x-ray 
diffraction measurement can be more fruitful by exploring 
lattice modulations generated by the underlaying CO. 
Given that the CO instability at $\vq \approx \vq_{1}$ is expected 
below $\delta \lesssim \delta_{c}$ \cite{bejas14}, it might seem easier to measure a sample 
with lower carrier density than 0.15. However, 
antiferromagnetism tends to be stabilized at lower 
carrier density and could mask the CO instability.

The peak positions at $\vq_{1}$ and $\vq_{2}$ found in $\chi_{d}(\vq,0)$ (fig. \ref{xq}) and 
$S(\vq)$ (fig. \ref{Sq}) are determined mainly by two factors: 
the so-called $2k_F$ scattering processes \cite{holder12} and the $d$-wave character of the bond order. 
The corresponding scattering processes are depicted in fig. \ref{fig:pattern}~(c). 
In particular, 
the peak at $\vq_{2}$ becomes pronounced substantially by the $d$-wave form factor. 
Hence we expect that the observed CO at $\vq=\vq_{2}$ \cite{da-dilva-neto15} 
has a $d$-wave character, which may be tested by RXS in the future\cite{comin15}.

As shown in fig. \ref{xqw}, charge excitations feature 
a V-shaped spectrum around $\vq=(0,0)$, which agrees qualitatively with the experimental 
observations \cite{wslee14,ishii14}. The spectra near $\vq=(0,0)$ come mainly  
from individual particle-hole excitations, in favor of the experimental 
interpretation by Ishii {\it et al.} \cite{ishii14}. 
A quantitative comparison 
with the experiment requires additional care. 
The V-shaped dispersion reported in 
\cite{ishii14} extends to 1.5 eV at $\vq=(0.6\pi,0)$ and $(0.6\pi,0.6\pi)$. Using 
$t/2=500$ meV \footnote{A factor of $1/2$ here comes from a large-$N$ formalism, where $t$ is scaled by $1/N$. 
We may then invoke $N=2$ when making a comparison with experiment.}, 
which is the estimated value for cuprates \cite{hybertsen90}, 
our obtained dispersion (fig. \ref{xqw}) extends up to 
$\omega\approx 500$ meV at the same momenta.  
This energy scale is about 
a factor of three lower than the experimental observation. 
Our small energy scale originates mainly from a relatively small band width due to the renormalization 
of the bare $t$ to an effective hopping $t_{\rm eff}=t \delta$, as seen in eq. \eq{eq:ek}. 
On the other hand, 
a large band width develops immediately after doping the Mott insulator phase, 
which cannot be captured quantitatively in terms of $t_{\rm eff}$. 
In this sense, a quantitative comparison of energy scale of charge excitations 
is connected with a fundamental issue of 
doped Mott insulators and remains to be studied. 
Furthermore in present study we focus on 
charge excitations of $d$-wave bond order because it gives 
the most relevant contributions to charge excitations at low energy. 
For a full comparison with RIXS data, however, 
charge excitations from other types of bond orders \cite{bejas14} as well as 
the usual charge density should be considered since 
RIXS may contain spectra of those excitations 
especially in a high energy region. 
At low energy, contributions from the $d$-wave bond order 
should become dominant and thus it is interesting to test a softening of charge 
excitation spectrum at $\vq_{1}$ and $\vq_{2}$ (see fig. \ref{xqw}) by RIXS and 
to clarify the actual energy scale there. 

The square lattice in the present model describes the Cu sites in the CuO$_{2}$ plane 
of cuprate superconductors and the center of the nearest-neighbor sites corresponds to  
the oxygen site. Therefore our bond order may be 
interpreted as a charge modulation at the oxygen sites \cite{comin15}. 

It is natural to consider whether the present theory can be applied also to hole-doped
cuprates by taking appropriate parameters. 
Comprehensive calculations in the hole-doped case \cite{bejas12}, however, did not 
capture a CO tendency compatible to the experimental observation 
\cite{wu11,ghiringhelli12, chang12, achkar12,leboeuf13,blackburn13,blanco-canosa14, comin14, 
da-silva-neto14, hashimoto14, tabis14}.  
Such calculations were performed in the paramagnetic state whereas in reality 
the CO is observed as an instability in the pseudogap state. 
Hence we consider that the effect of the pseudogap is crucial to understand the origin of 
the CO in hole-doped cuprates. In electron-doped cuprates, on the other hand, 
the CO tendency is observed in the paramagnetic state 
\footnote{Although a pseudogap  was reported in the optical conductivity spectra in the non-superconducting 
crystals \cite{onose01}, the pseudogap corresponding to the observed one in hole-doped cuprates, namely 
a gap-like feature above the superconducting phase,  is missing or at least very weak.}. 
This may be a reason why the present theory works for that case. 

While we have assumed the paramagnetic state, one may wonder 
whether the present theoretical framework actually predicts the strong asymmetry 
of the pseudogap between electron-doped cuprates ($t'>0$) and 
hole-doped cuprates ($t'<0$) simply by taking a different sign of $t'$. 
The origin of the pseudogap remains controversial even in hole-doped 
cuprates \cite{mishra14,hashimoto14b}.  If one assumes that the pseudogap 
is driven by a strong charge-order tendency as hinted in some 
experiments \cite{tanaka06,vishik10,kondo11,yoshida12},
a theoretical study \cite{bejas14} using the present large-$N$ scheme indeed suggests 
that pseudogap features should appear much weaker in electron-doped cuprates 
than hole-doped cuprates in the sense that a charge order tendency becomes much weaker 
in the former especially in a moderate doping region relevant to the pseudogap 
(see fig. 6 in \cite{bejas14}).

\section{Summary} 
Motivated by the recent measurements by RXS and RIXS in electron-doped cuprates, 
we have studied 
charge excitation spectra associated with 
a $d$-wave bond order 
in the two-dimensional $t$-$J$ model on a square lattice. 
We find that the static $d$-wave bond-order susceptibility $\chi_d(\vq, 0)$ has two peaks at 
$\vq_{1}=(0.84\pi, 0.84\pi)$ and $\vq_{2}=(0.49\pi,0)$, which are generated by the 
$2 k_{F}$ scattering processes enhanced by the $d$-wave character of bond-charge density. 
In spite of the proximity to the $d$-wave CO instability at  $\vq \approx \vq_{1}$, 
the peak at $\vq_{1}$ is very broad and becomes sharp only in the vicinity of its instability. 
On the other hand, the peak at $\vq_{2}$  
becomes sharper with decreasing temperature
but does not diverge, indicating that the CO with momentum $\vq_{2}$ is short ranged.  
These features are seen also in the equal-time correlation function $S(\vq)$. 
The spectral function of the $d$-wave bond order
(${\rm Im}\chi_{d}(\vq,\omega)$) forms a V-shape dispersion near $\vq=(0,0)$.   
This dispersion comes mainly from particle-hole excitations. 
The spectra bend back and reach close to zero energy at $\vq=\vq_{1}$ and $\vq_{2}$ 
where both the  static $d$-wave susceptibility and the  
equal-time correlation function show a peak. 
The resulting spectra have charge gap-like features 
with a maximal gap at $\vq\approx \frac{1}{2}\vq_{1}$ and $\frac{1}{2}\vq_{2}$.  
We argue that the CO observed in RXS is interpreted as a short-range order, 
which may not develop to become long range. 
It is interesting to explore gap-like features of the energy-resolved spectra 
and a possible CO near $\vq \approx \vq_{1}$ by RIXS and usual x-ray diffraction measurements, 
respectively, for electron-doped cuprates.

{\bf Acknowledgments}
The authors thank M. Fujita, K. Ishii, and T. Tohyama for stimulating discussions 
about RXS and RIXS for electron-doped cuprates. 
H.Y. acknowledges Instituto de F\'{\i}sica Rosario (UNR-CONICET) for hospitality 
and support by a Grant-in-Aid for Scientific Research from Monkasho. 

\bibliography{mainarxiv} 
\end{document}